\documentclass{article} 

\usepackage{iclr2027_conference,times}
\iclrfinalcopy 

\usepackage{amsmath,amsfonts,bm}

\def\eqref#1{equation~\ref{#1}}
\def\Eqref#1{Equation~\ref{#1}}

\def\1{\bm{1}}

\def\vc{{\bm{c}}}
\def\vd{{\bm{d}}}
\def\ve{{\bm{e}}}

\def\vh{{\bm{h}}}

\def\vm{{\bm{m}}}
\def\vn{{\bm{n}}}
\def\vo{{\bm{o}}}

\def\vs{{\bm{s}}}
\def\vt{{\bm{t}}}

\def\vv{{\bm{v}}}

\def\vx{{\bm{x}}}
\def\vy{{\bm{y}}}

\def\mO{{\bm{O}}}

\def\mX{{\bm{X}}}

\DeclareMathAlphabet{\mathsfit}{\encodingdefault}{\sfdefault}{m}{sl}
\SetMathAlphabet{\mathsfit}{bold}{\encodingdefault}{\sfdefault}{bx}{n}

\usepackage{microtype}
\usepackage{graphicx}
\usepackage{subcaption}
\usepackage{booktabs} 
\usepackage{mathtools}
\usepackage{amsmath}
\usepackage{siunitx}
\usepackage{xurl}
\usepackage{float}
\usepackage{colortbl}
\usepackage{tikz}
\usetikzlibrary{positioning}
\usetikzlibrary{shapes.geometric}
\usetikzlibrary{arrows.meta}
\usepackage{multicol}
\usepackage{amssymb}
\usepackage{pifont}
\usepackage{caption}

\usepackage{enumitem}
\usepackage{hyperref}
\usepackage{algorithm}
\definecolor{lightergray}{RGB}{230,230,230}
\definecolor{DarkGreen}{RGB}{30,130,30}

\usepackage{hyperref}
\usepackage{url}

\usepackage{xcolor}
\definecolor{QZeroFill}{HTML}{C8DCF4}
\definecolor{QOneFill}{HTML}{FBE8BF}
\definecolor{QTwoFill}{HTML}{F4C1D4}
\definecolor{QThreeFill}{HTML}{D0E8D0}
\definecolor{QFourFill}{HTML}{C6C1E3}
\usepackage[noend]{algorithmic}
\usepackage{hyperref}

\begin{document}

\title{AVSG: Accelerated Vectorized Sparse Gather for Efficient KV Cache Offload in Sparse-Attention LLM Serving}

\author{Wenwei Kuang\thanks{Corresponding author}, Xiangyu Wang, Chong Wu\footnotemark[1], Jun Wang\footnotemark[1],\\ \textbf{Weijie Zhang, Brian K Chen, Longwen Lan, Ken Zhang}\\Theory Lab, 2012 Labs, Huawei Technologies Co., Ltd.\\
\texttt{\{kuang.wenwei, wu.chong1, wang.jun7\}@huawei.com}
}

\maketitle

\begin{abstract}
Dynamic sparse attention reduces long-context attention computation by selecting only a subset of tokens, but still requires access to the full key-value (KV) cache, leaving serving memory-bound. Offloading the KV cache to host memory reduces device memory pressure but places host-to-device (H2D) transfers on the decoding critical path. In DeepSeek Sparse Attention (DSA), substantial overlap in selected KV entries across decoding steps creates an opportunity for device-resident reuse. Exploiting this reuse efficiently, however, presents three critical challenges: (1) costly matching of selected tokens against entries retained in the HBM buffer, (2) uneven distribution of the remaining H2D transfers across accelerator cores, and (3) retaining frequently accessed KV entries within limited HBM capacity. We present AVSG (Accelerated Vectorized Sparse Gather), an operator that addresses these challenges for DSA while preserving its exact token selections. AVSG uses vectorized hash matching to identify reusable HBM-buffer slots and reserve slots for entries requiring transfer, then evenly partitions these H2D transfers across accelerator cores. Lifetime-based buffer management retains frequently accessed KV entries in the device buffer across decoding steps, and shared slot reservations extend reuse across tokens within a multi-token prediction iteration. On a single NPU, vectorized hash matching is $2.80\times$ faster than scalar dual-pointer matching, miss-only transfer raises effective H2D bandwidth by up to $34.44\times$ over request-level assignment, and an 8K-entry buffer reaches a $94.83\%$ hit rate on real requests. These gains translate into end-to-end improvement: on a serving stack processing a real-world production dataset, AVSG reduces time per output token by $39\%$ and increases output throughput by $1.27\times$ relative to the same KV offload layout without HBM-buffer reuse. The code of AVSG is released at \url{https://gitcode.com/cann/cann-recipes-infer}.
\end{abstract}

\section{Introduction}
\label{sec:introduction}

Modern large language model (LLM) applications, including agentic tool use and retrieval-augmented assistants, increasingly operate on contexts approaching one million tokens, as exemplified by recent models such as GLM-5.3~\citep{glm53}, and DeepSeek-V4~\citep{deepseekv4}. For inference serving, the central objective is to sustain high aggregate throughput while controlling per-token latency, commonly measured by tokens per second (TPS) and time per output token (TPOT).

Serving such long contexts is fundamentally memory intensive. A request's key-value (KV) cache grows linearly with context length and can exceed the high-bandwidth memory (HBM) capacity of a single device~\citep{xu2026deepseek}.
Only a small number of long-context requests can therefore remain resident in HBM, limiting serving concurrency. DeepSeek Sparse Attention (DSA)~\citep{deepseek2025} reduces attention computation by selecting a fixed top-$K$ subset of historical tokens. However, because an indexer may select any earlier token, the full historical KV state and indexer-key state must remain addressable. Sparse computation alone does not eliminate the storage bottleneck.

KV cache offloading addresses this capacity problem by placing the full history in host memory and retaining only a working set on the accelerator. This frees HBM for additional requests, but places host-to-device (H2D) transfers on the decoding critical path. Sparse attention cannot proceed until the selected KV entries have been transferred to device memory. A naive implementation treats decoding steps independently and transfers all selected entries at every step.

Device-resident reuse pools (e.g., GatherSelection\footnote{\url{https://github.com/hicann/cann-recipes-infer/tree/master/ops/ascendc/src/gather_selection_kv_cache}}) and prefetching can reduce this cost but must address three coupled challenges. \textbf{First, matching overhead:} each selected token must be matched against the HBM buffer and mapped to a slot before attention; sequential lookup can offset transfer-time savings from reuse. \textbf{Second, fragmented transfers:} filtering out buffer hits leaves scattered KV entries for H2D transfer, with demands varying across requests and steps. Assignment by request or selected position can imbalance core workloads, limiting effective transfer bandwidth. \textbf{Third, effective HBM residency:} a limited buffer must retain frequently selected entries beyond the previous step while reclaiming less useful ones; repeatedly evicting useful entries wastes reuse opportunities and transfer bandwidth. Jointly addressing these challenges is necessary to translate reduced transfer volume into lower decoding latency.

\begin{figure}[t]
\centering
\includegraphics[width=0.87\textwidth]{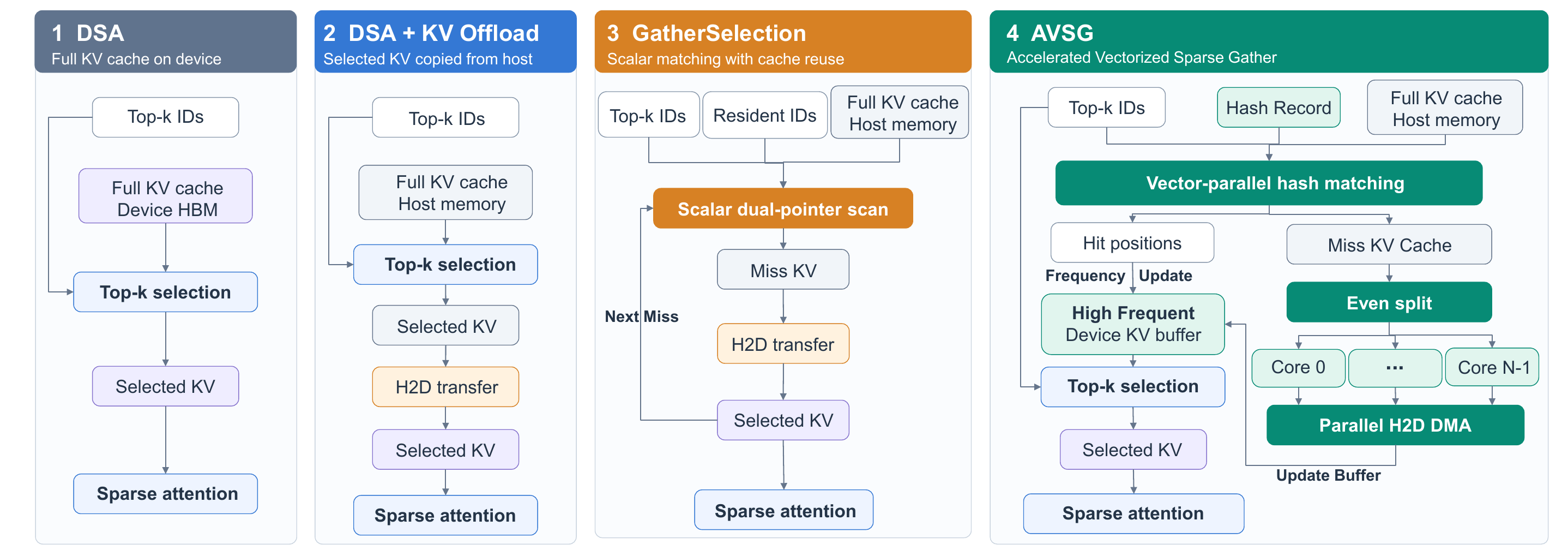}
\vspace{-3mm}
\caption{Evolution of KV transfer strategies for DSA inference. (1) HBM-resident full KV cache limits batch size. (2) KV offloading transfers all selected entries each step. (3) GatherSelection reuses a fixed-size cache through sequential matching. (4) AVSG reduces offloading latency through vectorized matching, balanced multi-core transfers, and retention of frequently accessed KV entries.}
\label{fig:architecture}
\vspace{-3mm}
\end{figure}

We present AVSG (Accelerated Vectorized Sparse Gather), an exact post-selection gather operator for offloaded DSA inference (Figure~\ref{fig:architecture}-4). AVSG jointly addresses these challenges by reducing redundant KV transfers and accelerating the matching and movement of selected entries, without modifying the indexer or its token selections. \textbf{Our contributions} are as follows:
\begin{itemize}
    \item \textbf{Vectorized Hash Matching via Open Addressing}: We replace scalar dual-pointer matching with vectorized hash probing, decoupling device buffer capacity $S$ from selection size $K$ without a proportional matching penalty. Matching is $2.80\times$ faster at $S=K=2048$, while latency grows sublinearly with $K$ and $S$ over the tested ranges (\S\ref{subsec:hash-matching}--\ref{subsec:stage-one}; \S\ref{sec:hashmatch}).
    \item \textbf{Evenly Partitioned Multi-Core H2D Transfer}: We balance cache-miss KV transfers across AI-Cores to reduce transfer imbalance, achieving up to $34.44\times$ and $3.48\times$ the effective H2D bandwidth of request-level assignment and equal top-$K$ partitioning (\S\ref{subsec:two-stage}; \S\ref{subsec:h2d-efficiency}).
    \item \textbf{Device Cache for Frequently Accessed KV Entries}: We combine per-slot aging with hit-triggered lifetime refresh to retain frequently accessed KV entries across decoding steps and reclaim inactive slots. On the evaluated real requests, an 8K-entry buffer achieves a $94.83\%$ hit rate, $13.22\%$ above the adjacent-step reuse rate (\S\ref{subsec:cache-buffer}; \S\ref{subsec:cache-reuse}).
    \item \textbf{Shared-Buffer Reuse Across MTP Tokens}: We sequentially process the tokens of a multi-token prediction (MTP) iteration through a shared buffer, allowing later tokens to reuse earlier slot reservations. This achieves $61.87\%$ reuse in the first decode iteration from an empty cache under MTP-3 (\S\ref{subsec:mtp}; \S\ref{subsec:cache-reuse}).
\end{itemize}
Together, these mechanisms reduce decode TPOT by $39\%$ and increase throughput by $1.27\times$ over the same KV offload layout without HBM-buffer reuse during end-to-end serving evaluation (\S\ref{subsec:e2e}).
\section{Background and Related Work}
\label{sec:related}
\paragraph{LLM Inference and KV Cache Management.}
PagedAttention reduces KV fragmentation through paged allocation, while RadixAttention reuses prefixes across requests through a radix tree~\citep{kwon2023efficient,SGLang}. Neither reduces an active request's KV storage requirement, so offloading remains necessary when aggregate KV demand exceeds HBM capacity.
\paragraph{Sparse Attention Mechanisms.}
Sparse attention reduces computation through reformulation or token selection~\citep{CURSA}. Training-free methods use structured patterns or approximate selection without retraining, at potential accuracy cost~\citep{xiao2024streamingllm,jiang2024minference,tang2024quest,lee2024infinigen,NEURIPS2025_3ab86803,ELFATT,BFLA}; training-time methods incorporate sparsity during training, including MoBA, NSA, InfLLM-v2, and DSA~\citep{lu2025moba,yuan2025nsa,zhao2025infllmv2,deepseek2025}. DSA uses a lightweight learned indexer to select top-$K$ tokens for sparse MLA, bounding per-operation KV reads while full KV and indexer-key storage still grow with context length. IndexCache~\citep{glm52} eliminates up to 75\% of per-layer indexer computations through cross-layer reuse; this is orthogonal to AVSG's acceleration of post-selection KV materialization.
\paragraph{KV cache offloading.}
KV-cache offloading extends the storage hierarchy beyond accelerator HBM to support long-context execution, cross-request reuse, and disaggregated serving. LMCache~\citep{cheng2025lmcache} offloads KV states to hierarchical storage for cross-query reuse; CachedAttention~\citep{gao2024cachedattention} organizes host-memory/disk tiers for cross-turn reuse with overlapped access; Mooncake~\citep{qin2025mooncake} pools CPU, DRAM, SSD, and NIC resources into a distributed cache within a prefill--decode disaggregated system; Strata~\citep{xie2025strata} combines GPU-assisted KV-cache I/O with cache-aware scheduling; and Pensieve~\citep{yu2025pensieve} maintains conversational state across requests with a two-tier GPU--CPU cache. These systems target storage capacity, request-level reuse, and cache placement---not the per-operation selection of KV entries that sparse attention requires.
\paragraph{KV offloading acceleration.}
Fine-grained, sparse-attention-oriented systems instead reduce the cost of materializing the KV entries each attention operation needs. ArkVale~\citep{chen2024arkvale} evicts and recalls KV pages by query-dependent importance; InfiniGen~\citep{lee2024infinigen} prefetches via next-layer attention rehearsal; FreeKV~\citep{liu2025freekv} and LiteCache~\citep{yi2025clo} reuse previously selected KV across adjacent decoding steps with speculative correction; SparseServe~\citep{zhou2025sparseserve} adds fragmentation-aware transfers and working-set-aware batching; ECHO~\citep{liu2026echo} hides offload latency with lossless intra/inter-query prefetching fused with indexer computation. GatherSelection (Figure~\ref{fig:architecture}(3)) belongs to this family but starts from the exact Top-$K$ identifiers: it sort--merge matches requested against cached token ids, retains same-row hits, and fetches only misses from the full KV cache. AVSG preserves this exact post-selection interface but replaces repeated sort--merge matching with a persistent hash-and-lifetime pool with explicit slot remapping, combining vectorized hash probing, fine-grained core allocation, and overlapped remapping and H2D transfer to retain frequently accessed KV entries across steps without modifying the indexer.

\begin{figure}[t]
\centering
\includegraphics[width=0.97\textwidth]{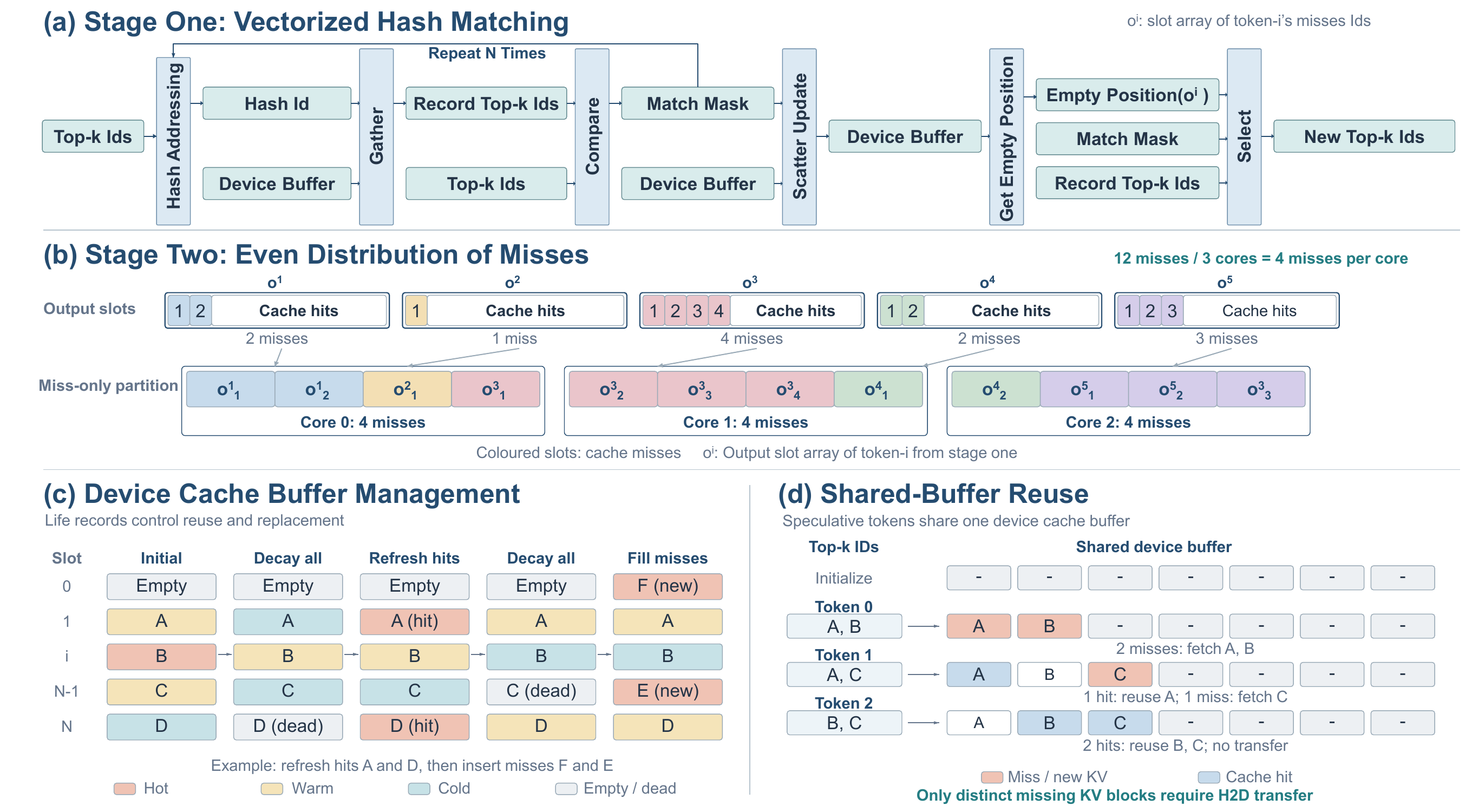}
\caption{Detailed AVSG algorithm design. \textbf{(a)} Vectorized open-addressing matching and updates of top-$K$ IDs. \textbf{(b)} Even distribution of KV misses across AICores for efficient H2D transfer. \textbf{(c)} Retention of frequently accessed KV entries and replacement of cold entries. \textbf{(d)} A shared buffer unions top-$K$ selections across a request's MTP tokens to avoid duplicate copies.}
\label{fig:avsg-detail}
\end{figure}

\section{Method}
\label{sec:method}
\subsection{Overview}
\label{subsec:overview}
We distinguish the full host-resident {KV cache} from AVSG's \emph{HBM buffer}, which holds selected KV entries and their reserved destinations. Unless otherwise stated, a buffer hit means matching finds a previously reserved, reusable slot; a buffer miss requires a new slot reservation and an H2D copy. These terms describe reuse in the HBM buffer, not whether the KV entry exists in the full KV cache.

Figure~\ref{fig:avsg-detail} illustrates the AVSG operator. At each decode step DSA supplies a set of $K$ top-$K$ token indices, and AVSG must make the corresponding KVs resident in device memory before the attention operator reads them. Because the full KV cache remains in host memory, the quantity AVSG controls is how much of it crosses the H2D link: instead of fetching every selected index, AVSG carries a device cache buffer of $S$ slots across steps and transfers only indices not already resident. Within a step, AVSG resolves every index against the buffer (Figure~\ref{fig:avsg-detail}a), classifying it as hit or miss and assigning each miss a destination slot, then fetches exactly the misses from host memory spread across the available AICores (Figure~\ref{fig:avsg-detail}b).

What makes the hit/miss split worth exploiting is the policy deciding which indices stay resident. Each slot carries a life record that decays every step, is refreshed on hit, and marks the slot reusable once expired, so the buffer keeps the indices DSA selects repeatedly and releases the rest (Figure~\ref{fig:avsg-detail}c). Let $N_{\mathrm{mtp}}$ denote the number of additional speculative tokens per request. The initial token and these $N_{\mathrm{mtp}}$ speculative tokens share a single buffer and are processed one after another (Figure~\ref{fig:avsg-detail}d), so indices one token fetches as misses are hits for the tokens behind it.


\subsection{Hash-Stat e: A Device-Resident Selection Pool}
\label{subsec:hash-matching}
To map selected token IDs to reusable buffer slots and allocate destinations for misses, AVSG organizes its metadata in a \emph{Hash-State} table.
The table resides entirely in on-chip memory and supports SIMD~\cite{flynn1972computer} instructions, like probing, comparison, and updates.

\textbf{Problem formulation.} At each decode step, each token contributes $K$ selected indices $\vx=(x_0,\ldots,x_{K-1})$ (e.g., $K{=}2048$). Each request contributes $N_q=N_{\mathrm{mtp}}+1$ tokens. A batch of $B$ requests contains $N_g=B\times N_q$ tokens. Tokens within a request share an $S$ slot hash table. Each table retains KV address fetched on previous steps until its life records expire (\S\ref{subsec:cache-buffer}), allowing repeated selections to reuse cached entries. Given $\vx$ and the request's table, Stage One maps each selected index $x_i$ ($i \in \{0,\ldots,K-1\}$) to a hit indicator $m_i\in\{0,1\}$ and a KV slot $o_i\in\{0,\ldots,S-1\}$:
\begin{equation}
m_i=\begin{cases}1, & \text{if } x_i \text{ is resident in the table},\\ 0, & \text{otherwise (a cache miss)},\end{cases}
\label{eq:hitmiss}
\end{equation}
where $o_i$ names the slot that already holds the token (hit) or is freshly allocated to hold it (miss); misses are fetched from host memory in Stage Two. Probing is governed by three parameters: two distinct primes $a_1,a_2$ that set a probe's initial bucket and stride, a retry budget $R$ bounding the probe chain (default $R{=}3$, i.e., $R{+}1$ candidate buckets per token), and the maximum lifetime $L$, to which a hit resets the slot's lifetime (default $L{=}16$, chosen via the ablation in Appendix~\ref{app:lifetime}).

\textbf{Hash-State table.}
To support these probes and lifetime updates, AVSG maintains a \emph{Hash-State} table $\vs=(E_0,\ldots,E_{S-1})$ of 64-bit entries. Each entry $E_j=(c_j,q_j)$ stores a full 32-bit selected index $c_j$ and a 32-bit metadata word $q_j$, partitioned as
\begin{equation}
q_j = \Bigl[
\underbrace{\text{lifetime}\vphantom{p}}_{8\,\text{bits}}
\;\Big|\;
\underbrace{\text{invalid}\vphantom{p}}_{1\,\text{bit}}
\;\Big|\;
\underbrace{\text{id\_footprinter}\vphantom{p}}_{(23-m)\,\text{bits}}
\;\Big|\;
\underbrace{\text{slot}\vphantom{p}}_{m\,\text{bits}}
\Bigr].
\label{eq:packed-hash-entry}
\end{equation}
The capacity is a power of two, $S{=}2^m$, with $m$ the slot bit width, so one packed word can identify any slot (e.g., $m{=}13$ gives $S{=}8192$).
In \Eqref{eq:packed-hash-entry}, fields are listed from the most significant bit to the least, so slot occupies the low $m$ bits of $q_j$.
$c_j$ stores the selected indice occupying slot $j$, while $q_j$'s lifetime tracks its access recency for eviction. 
Hash probing visits bucket $h_i$ for $x_i$; following the slot field $o_i$ of $q_{h_i}$ retrieves $c_{o_i}$, which is compared with $x_i$ to determine $m_i$.
The invalid and id\_footprinter fields of $q_{h_i}$ is used to determine whether $o_i$ can be overwritten during allocation (\S\ref{subsec:stage-one}).

This Hash-State layout enables vectorized parallel access to the table in an AICore's unified buffer (UB) using SIMD instructions. Section~\ref{subsec:stage-one} describes how these instructions implement parallel matching and hash-table state updates.

\textbf{Notation convention.} Boldface symbols such as $\vm,\vo,\vh,\vd,\vc$ denote the length-$K$ vector counterparts of the scalars $m_i,o_i,h_i,d_i,c_i$, and every operation on them executes as one SIMD instruction. A \emph{masked assignment} $\vy[\vm]\gets\vt$ updates only the elements of $\vy$ whose mask $\vm_i{=}1$, leaving the rest unchanged; $\vy[\neg\vm]$ updates the elements with $\vm_i{=}0$.

\subsection{Stage One: Three-Pass Vectorized Matching}
\label{subsec:stage-one}
Stage One turns the hit/miss classification of Eq.~\ref{eq:hitmiss} into three passes over the UB-resident table, each with a distinct responsibility (Algorithm~\ref{alg:avsg-pipeline}): a \emph{refresh pass} that maintains recency, a \emph{classification pass} that produces the outputs Stage Two consumes, and an \emph{allocation pass} that installs the misses into the table.
The refresh pass decays all lifetimes by one only for the first token of each request, then calls Probe for the current token and restores the lifetimes of matched entries to $L$ before allocating misses.
The classification pass reuses the hit mask $\vm$ and bucket indices $\vh$ from the refresh pass to record hit slots $o_i$ and compute the token's miss count $n_g$. Allocation fills remaining slots for misses, yielding device positions for downstream attention without a separate gather kernel. The per-token miss counts, collected across the batch as $\vn=(n_0,\ldots,n_{N_g-1})$, drive Stage Two's load balancing (\S\ref{subsec:two-stage}).
The allocation pass frees capacity---collecting dead slots and, when fewer than the miss count, decaying all records further until enough die, the vectorized analogue of evicting the $n_g$ least-recent entries---then installs each miss into a freed position and links a hash bucket to that slot.

\textbf{Probe primitive.}
The refresh pass calls the read-only primitive $\mathrm{Probe}(\vx,\vs)$ once per token, returning the hit mask $\vm$ and bucket-index vector $\vh$ (Appendix~\ref{app:alg-details}, Eq.~\ref{eq:probe-outputs}): given a token's top-$K$ list $\vx$, compute initial buckets $\vh \gets (a_1\vx)\bmod S$ and odd strides $\vd \gets ((a_2\vx)\bmod S)\,|\,1$, then repeat up to $R$ times---gather each bucket's metadata word, follow its slot, gather the cached id $\vc$, and advance unresolved probes by their strides---with one final un-advanced comparison yielding the hit mask $\vm \gets [\vc = \vx]$, so each $x_i$ examines at most $R{+}1$ buckets.
Every probe runs the full vector of ids, costing $O(1)$ SIMD instructions regardless of $K$ and $S$; this constant-time probing is what lets the hash table be scaled to raise the hit rate without a proportional matching penalty.

\textbf{Allocation mechanics.}
Mapping index $x_i$ to slot $o_i$ requires checking whether $q_{h_i}$ can be overwritten: its slot field may still point to a slot whose previous occupant has been replaced.
AVSG uses fingerprint verification and lazy reclamation
to identify and overwrite such stale mappings.
Using the fields of $q_{h_i}$ defined in \S\ref{subsec:hash-matching}, AVSG verifies the mapping by comparing the stored fingerprint with that of $c_{o_i}$.
Since bit 23 of the 24-bit index field is reserved as the invalid flag, a top-$K$ id $x$ has $F=2^{23}/S=2^{23-m}$ representable fingerprints, encoded element-wise as $\phi(x)=x\bmod F$ and $\widehat{\phi}(x)=\phi(x)\,S$; the packed word
\begin{equation}
q_{h_i} = 2FS\,\ell_{h_i} + FS\,b_{h_i} + \widehat{\phi}(c_{o_i}) + {o_i}
\label{eq:packed-hash-mapping}
\end{equation}
places the preserved lifetime $\ell_{h_i}$, the invalid flag $b_{h_i}$, the stored fingerprint, and the slot $o_i$ in the bit ranges declared in \Eqref{eq:packed-hash-entry}, so summation is equivalent to concatenation.
Stale pointers are reclaimed lazily: because AVSG keeps no reverse index from a slot back to its pointing buckets, an eager repair would require a full $O(S)$ scan, so staleness is detected at bucket-reuse time, riding on the metadata read that must happen anyway before an overwrite---a bucket is reusable if its mapping is invalid, if the pointed slot is inactive, or if the fingerprint no longer matches the slot's occupant,
\begin{equation}
\operatorname{reusable}(q_{h_i})
=
\neg\,\operatorname{valid}(q_{h_i})
\;\lor\;
\neg\,\operatorname{active}\bigl(q_{o_i}.\text{lifetime}\bigr)
\;\lor\;
\bigl[\phi_{h_i}\neq\phi(c_{o_i})\bigr],
\label{eq:reusable-hash-bucket}
\end{equation}
where $\operatorname{valid}(q_{h_i})=[b_{h_i}=0]$ and $\operatorname{active}(\ell)=[\ell\ge0]$. Here $\phi_{h_i}$ denotes the stored fingerprint, and $o_i=q_{h_i}\bmod S$ extracts the slot field defined in \S\ref{subsec:hash-matching}; the full decode chain appears in Appendix~\ref{app:decode-chain}.
An undetected stale pointer may increase probing overhead, but full 32-bit index comparisons prevent false hits (\S\ref{subsec:hash-matching}). The allocation pass applies the reusable predicate to all misses in vectorized form (Algorithm~\ref{alg:avsg-UpdateState}, Appendix~\ref{app:alg-details}), using the same initial buckets and odd strides. Every miss writes its selected index and initializes the lifetime in its allocated slot; only probes that find reusable buckets install the corresponding mappings.

\begin{algorithm}[t] \small
\caption{Stage One: Three-Pass Vectorized Matching Per Token}
\label{alg:avsg-pipeline}
\begin{algorithmic}[1]
\REQUIRE $g\in\{0,\ldots,N_q-1\}$: token index within the request;\quad selected indices $\vx$ and request-specific table $\vs$
\ENSURE Slot vector $\vo$;\quad per-token miss count $n_g$;\quad updated table $\vs$
\STATE \texttt{\#}\ \textbf{Pass 1 --- Refresh}: decay, then restore records of hits (Steps 1--2 of Algorithm~\ref{alg:avsg-lifecycle})
\IF{$g=0$}
    \STATE $\vs.\text{lifetime} \gets \vs.\text{lifetime} - 1$
\ENDIF
\STATE $\vm, \vh \gets \mathrm{Probe}(\vx,\, \vs)$ \COMMENT{hit mask $\vm$, bucket indices $\vh$; Eq.~\ref{eq:probe-outputs}}
\STATE $\vo^{\mathrm{match}} \gets q_{\vh[\vm]}.\text{slot}$
\STATE $q_{\vo^{\mathrm{match}}}.\text{lifetime} \gets L$
\STATE \texttt{\#}\ \textbf{Pass 2 --- Classify}: emit slot indices and per-token miss counts.
\STATE $\vo[\vm] \gets \vo^{\mathrm{match}}$
\STATE $n_g \gets \sum_{i=0}^{K-1}(1-m_i)$ \COMMENT{$m_i$ is element $i$ of the hit mask $\vm$}
\STATE \texttt{\#}\ \textbf{Pass 3 --- Allocate}: free slots, install misses, link buckets (Steps 3--4 of Algorithm~\ref{alg:avsg-lifecycle})
\STATE $\ve \gets \mathrm{CollectDead}(\vs)$ \COMMENT{Eq.~\ref{eq:collect-dead}, Appendix~\ref{app:alg-details}}
\WHILE{$|\ve| < n_g$}
    \STATE $\vs.\text{lifetime} \gets \vs.\text{lifetime} - 1$;\quad $\ve \gets \mathrm{CollectDead}(\vs)$
\ENDWHILE
\STATE $\vt \gets \vx[\neg\,\vm]$;\quad $\vo^{\mathrm{miss}} \gets \ve[:n_g]$ \COMMENT{selected indices that miss take freed slots}
\STATE $\vs \gets \mathrm{UpdateState}(\vt,\, \vo^{\mathrm{miss}},\, \vs)$ \COMMENT{Algorithm~\ref{alg:avsg-UpdateState}}
\STATE $\vo[\neg\,\vm] \gets \vo^{\mathrm{miss}}$ \COMMENT{fill placeholders}
\end{algorithmic}
\end{algorithm}

\subsection{Fine-Grained Multi-Core H2D Transfer}
\label{subsec:two-stage}

After matching, each selected index $x_i$ with $m_i=0$ requires an H2D transfer from position $x_i$ in the host KV cache to slot $o_i$ in the device buffer. Each AICore performs these transfers through its Direct Memory Access (DMA) engine,reading host memory directly without an intermediate staging copy. Under DSA's single-MLA-head configuration, a batch contains $N_g=B\times N_q$ tokens, each contributing $K$ selected indices and their matching results, which together determine the batch's H2D transfer workload.

With $C$ AICores, let $W_c$ denote the H2D transfer workload of core $c\in\{0,\ldots,C-1\}$. $\mathcal A_c$ is defined as the set of selected-index positions $(g,i)$ assigned to core $c$, where $g$ identifies a token and $i$ a position in its selected indices. Since only misses require transfer, each position contributes $1-m_{g,i}$ KV entries, where $m_{g,i}$ is its hit indicator. The resulting transfer workload is
\begin{equation}
W_c=\sum_{(g,i)\in\mathcal A_c}(1-m_{g,i})
=|\mathcal A_c|-\sum_{(g,i)\in\mathcal A_c}m_{g,i}.
\label{eq:core-transfer-workload}
\end{equation}

AVSG constructs each core's assignment $\mathcal A_c$ using only misses, so $m_{g,i}=0$ for all $(g,i)\in\mathcal A_c$. Consequently, $\sum_{(g,i)\in\mathcal A_c}m_{g,i}=0$, yielding $W_c=|\mathcal A_c|$ from Eq.~\ref{eq:core-transfer-workload}. For token $g$, the miss count is $n_g=\sum_{i=0}^{K-1}(1-m_{g,i})$, giving $N_{\text{miss}}=\sum_{g=0}^{N_g-1}n_g$ transfers across the batch.
AVSG orders these misses by token and partitions them into $C$ disjoint assignments covering every miss exactly once. Each core receives either $\lfloor N_{\text{miss}}/C\rfloor$ or $\lceil N_{\text{miss}}/C\rceil$ consecutive misses. For any two cores $c,d\in\{0,\ldots,C-1\}$,
\begin{equation}
W_c=|\mathcal A_c|\in
\left\{\left\lfloor\frac{N_{\text{miss}}}{C}\right\rfloor,
\left\lceil\frac{N_{\text{miss}}}{C}\right\rceil\right\},
\qquad |W_c-W_d|\le1.
\label{eq:balanced-transfer-workload}
\end{equation}
Thus, the H2D workloads of any two cores differ by at most one KV entry, regardless of the original request or cache-hit distribution.
Section~\ref{subsec:h2d-efficiency} compares AVSG with request-level assignment and equal top-$K$ partitioning; Appendix~\ref{app:h2d-details} analyzes the transfer imbalance of these baselines.

\subsection{Device Cache for Frequently Accessed KV Entries}
\label{subsec:cache-buffer}
Within a request, $N_q$ tokens, each with $K$ selected indices,  share an $S$-slot hash table. To accommodate all corresponding slots even when the $N_qK$ indices are distinct, the capacity must satisfy
\begin{equation}
S \;\geq\; N_qK = (N_{\mathrm{mtp}} + 1)K,
\label{eq:buffer-capacity}
\end{equation}
ensuring enough capacity for the request's full working set even if every selection misses, without evicting entries still needed in the current step.

\textbf{Lifetime-Based Slot Management.}
The lifetime field of $q_j$ carries an 8-bit counter ($0$--$L$) that tracks access recency.
The updates in Algorithm~\ref{alg:avsg-pipeline} ensure that unmatched old slots expire before slots matched by the current request.
Global pre-decay leaves unmatched old slots with life at most $L-1$, while the request-wide refresh restores every matched slot to $L$.
Together with the capacity bound in Eq.~\ref{eq:buffer-capacity}, this preserves the slots required by the current request. The policy implements LRU-like aging based on access recency, rather than an exact LRU ordering.
Appendix~\ref{app:alg-details} provides the consolidated update procedure and a proof of this preservation guarantee.

\textbf{Frequently Accessed Data Residency.}
Because the resident set is determined by access recency, the hash table content tracks the workload's top-K selection pattern.
For each selected index $x_i$, the lifetime associated with
$c_{o_i}$ is set to $L$, either by refreshing a hit or initializing
a newly allocated entry. Without another refresh, the slot
becomes reclaimable after more than $L$ unit decays.
Repeated selections thus prolong residency, while unaccessed
entries eventually expire.

\subsection{Shared-Buffer Reuse Across MTP Tokens}
\label{subsec:mtp}

AVSG supports shared-buffer reuse across MTP tokens through two mechanisms: reuse of matching information to avoid redundant KV transfers within a step, and speculative-entry invalidation to preserve correctness across steps.

\textbf{Progressive Reuse of Matching Information.}
A later token can match a slot reserved by an earlier token even before that slot's KV data arrive: both tokens refer to the same destination slot, and the corresponding miss requires only one H2D copy. At cold start the initial token reserves slots for its misses, while later tokens reuse those reservations and reduce the total transfer volume. We evaluate the resulting first-step reuse in \S\ref{subsec:cache-reuse}.

\textbf{Speculative-Entry Invalidation.}
For a selected index $x_i$, the slot field of $q_{h_i}$ identifies $o_i$, and (\S\ref{subsec:two-stage}) copies the corresponding host KV entry into device slot $o_i$.
If $x_i$ is a speculative index, its host KV entry may be overwritten during subsequent decoding, leaving the device copy at $o_i$ stale. Retaining the mapping could then yield a hit ($m_i=1$) based on the unchanged index and reuse KV data that no longer match the host entry, compromising inference accuracy. AVSG therefore marks the selected index in the referenced slot as empty ($c_{o_i}\gets-1$) for each speculative index. Subsequent comparisons with $x_i$ then fail, preventing stale hits and requiring a fresh H2D copy when the index is reselected.

\section{Evaluation}
\label{sec:evaluation}

\subsection{Experimental Setup}
We compare three configurations under identical request workloads: \textbf{Baseline (No Offload)}, which keeps all KV in HBM with device capacity capped by the memory left after model weights; \textbf{KV Offload}, which offloads the full KV cache to host memory and transfers every activated KV at each decode step with no reuse; and \textbf{AVSG}, which augments the same offload layout with a cache buffer that retains frequently accessed KV entries, vectorized hash matching, and fine-grained two-stage transfer.
The component-level experiments (\S\ref{sec:hashmatch}--\S\ref{subsec:cache-reuse}) isolate each mechanism on a single NPU; the end-to-end serving comparison (\S\ref{subsec:e2e}) runs the full 5P1D serving stack, whose platform and memory layout we describe there.

\subsection{Vectorized Hash Matching Performance}
\label{sec:hashmatch}

\begin{figure}[t]
\centering
\includegraphics[width=0.9\linewidth]{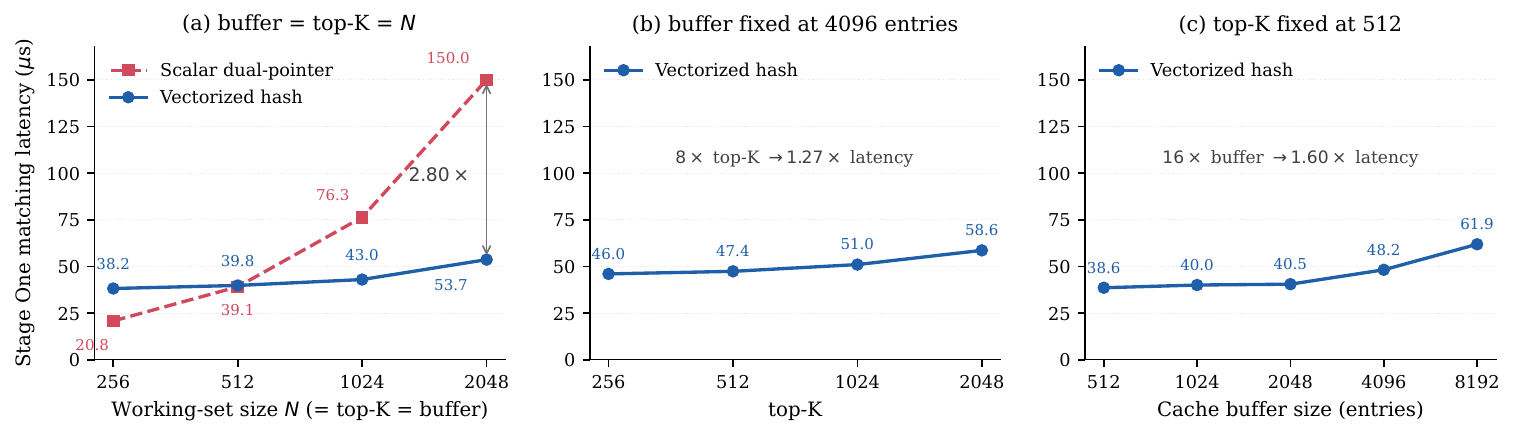}
\caption{Stage One matching latency (median of 1000 iterations). \textbf{(a)} Scalar versus vectorized matching at $S=K=N$. \textbf{(b)} Varying $K$ at $S=4096$. \textbf{(c)} Varying $S$ at $K=512$.}
\label{fig:hashmatch}
\end{figure}

Vectorized hash matching reduces lookup latency while allowing the device buffer to grow without a proportional matching penalty, supporting the design in \S\ref{subsec:stage-one}. Figure~\ref{fig:hashmatch} shows a reduction from $150.0$ to $53.7\,\mu$s at $K=S=2048$ relative to the scalar dual-pointer baseline ($64.2\%$ lower latency, $2.80\times$ faster). Increasing the selection size $K$ by $8\times$ at fixed buffer capacity $S$ raises latency by only $1.27\times$; increasing $S$ by $16\times$ at fixed $K$ costs $1.60\times$. This sublinear growth over the tested ranges makes it practical to retain more reusable KV entries without a comparable increase in lookup time. Appendix~\ref{app:hashmatch-details} provides the profiling protocol and detailed analysis.

\subsection{Fine-Grained Multi-Core H2D Transfer Efficiency}
\label{subsec:h2d-efficiency}

\begin{figure}[H]
\centering
\includegraphics[width=0.85\linewidth]{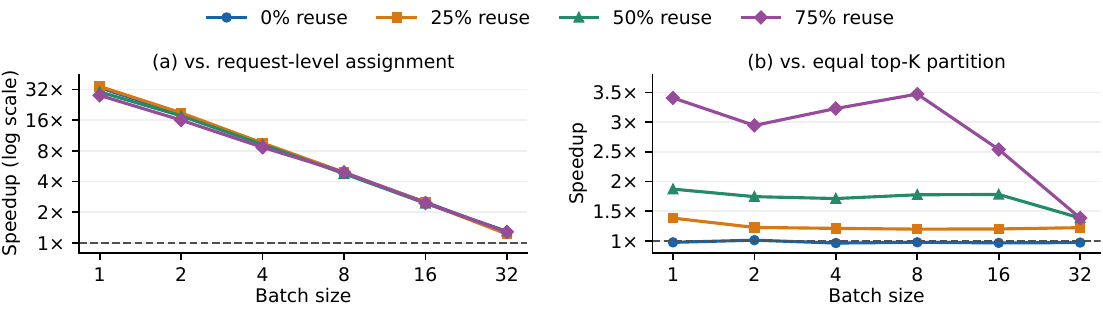}
\caption{H2D bandwidth gains from evenly splitting cache misses across AI-Cores: \textbf{(a)} over request-level assignment, which leaves cores idle at small batches; \textbf{(b)} over equal top-$K$ partitions, whose transfer workloads can become uneven after cache reuse.}
\label{fig:h2d-speedup}
\end{figure}

Evenly distributing actual cache misses across AI-Cores improves H2D efficiency both when the batch is small and cache reuse leaves fewer KVs to transfer, as intended by \S\ref{subsec:two-stage}. In Figure~\ref{fig:h2d-speedup}(a), AVSG achieves up to $34.44\times$ the bandwidth of request-level assignment ($B=1$, $25\%$ reuse), demonstrating the benefit of spreading a request's transfers across cores. Against equal top-$K$ partitions, it achieves up to $3.48\times$ higher bandwidth ($B=8$, $75\%$ reuse; Figure~\ref{fig:h2d-speedup}(b)), supporting the choice to balance the remaining copies after identifying cache hits. Full measurements and analysis appear in Appendix~\ref{app:h2d-details} (Table~\ref{tab:h2d-full-speedup}).

\subsection{Cache Reuse Rate Analysis}
\label{subsec:cache-reuse}

AVSG reduces repeated KV copies by retaining frequently accessed entries across decode steps and sharing slot reservations within each MTP iteration (\S\ref{subsec:cache-buffer}, \S\ref{subsec:mtp}). On the 2000 requests of \S\ref{subsec:e2e} with $K=2048$, adjacent-step reuse---the fraction of selected entries also selected in the preceding step---is $81.61\%$. The 4K and 8K buffers achieve hit rates of $88.72\%$ and $94.83\%$, respectively, showing that additional capacity preserves useful entries beyond the previous step. Table~\ref{tab:reuse-summary} also reports the best synthetic gains at $50\%$, $70\%$, and $90\%$ adjacent-step reuse. Even from an empty buffer, MTP-3 enables four tokens (one initial and three speculative) to achieve $61.87\%$ reuse, reducing average KV copies from 8192 to $8192(1-61.87\%)\approx3124$. The initial token still misses; later tokens reuse its slot reservations before the collective H2D transfer. Appendix~\ref{app:reuse-details} examines capacity across reuse distributions and cold-start reuse across MTP degrees, while Appendix~\ref{app:lifetime} evaluates lifetime versus reuse and kernel execution time.

\begin{table}[t] \small
\centering
\scriptsize
\caption{Cache reuse ($K=2048$). Rates are percentages; gain is 8K hit rate minus adjacent-step reuse (pp). Synthetic rows maximize this gain at each reuse level, all with historical recall mean $0.9$.}
\vspace{-3mm}
\label{tab:reuse-summary}
\begin{tabular}{@{}lrrrr@{}}
\toprule
Data & Adjacent-step reuse & 4K hit rate & 8K hit rate & Gain (pp) \\
\midrule
Synthetic & 50.00 & 84.29 & 94.15 & +44.15 \\
Synthetic & 70.00 & 90.12 & 96.79 & +26.79 \\
Synthetic & 90.00 & 96.21 & 99.04 & +9.04 \\
Real & 81.61 & 88.72 & 94.83 & +13.22 \\
\midrule
\multicolumn{5}{@{}l}{\textbf{MTP-3 cold start ($S=8$K, four tokens per request)}} \\
Metric & \multicolumn{2}{c}{Independent buffers} & \multicolumn{2}{c@{}}{Shared buffer} \\
\midrule
Cache hit rate (\%) & \multicolumn{2}{c}{0} & \multicolumn{2}{c@{}}{61.87} \\
KV entries copied (avg.) & \multicolumn{2}{c}{8192} & \multicolumn{2}{c@{}}{$3124$} \\
\bottomrule
\end{tabular}
\end{table}
\subsection{End-to-End Serving Performance}
\label{subsec:e2e}

The component experiments above validate each mechanism in isolation; this subsection asks whether the gains survive integration into a full serving stack. All three settings run a 5P1D dis-aggregated topology (12 nodes of 910B NPUs) serving GLM-5.3-w4a8c16 with 3-token speculative decoding. The workload is an agent dataset with a total of 2000 real chat requests collected from a production system. The average prompt length is 48,000 tokens, and we test with a maximum output of 32768 tokens, a concurrency of 64, and a request rate of 1. We extend the closed-loop benchmarking methodology of \citet{wang2025omniinfer}: throughout the evaluation the system is bounded either by fixed in-flight concurrency or by the request injection rate, so it is continuously stressed up to its throughput limit. In this regime, higher throughput naturally comes with a larger TTFT, as more concurrently served requests contend for prefill compute, so the two metrics trade off and must be read together.

\begin{table}[h]
\centering \small
\caption{End-to-end cold-start comparison (5P1D, GLM-5.3-w4a8c16). TPOT from the second token onward; HBM usage is the per-chip decode average.}
\vspace{-3mm}
\label{tab:e2e}
\scriptsize
\begin{tabular}{lccc}
\toprule
\textbf{Metric} & \textbf{No Offload} & \textbf{KV Offload} & \textbf{AVSG} \\
\midrule
Device KV capacity (tokens) & 145{,}536 & 762{,}496 & 762{,}496 \\
Decode HBM usage (avg) & 91.8\% & 91.6\% & 95.5\%  \\
AVG\_TTFT (s) & 7.92 & 5.94 & 8.69 \\
AVG\_TPOT (ms) & 43 & 91 & \textbf{55} \\
TP99\_TPOT (ms) & 64 & 126 & \textbf{77} \\
AVG\_E2E (s) & 119.1 & 120.2 & \textbf{75.8} \\
Output TPS (tok/s) & 472.6 & 451.2 & \textbf{571.4} \\
Achieved request rate (req/s) & 0.44 & 0.40 & \textbf{0.54} \\
\bottomrule
\end{tabular}
\end{table}

No Offload attains the lowest TPOT (43~ms, no H2D on the decode path) but is capacity-capped: its 145K-token device pool cannot sustain the 48K-token prompts at concurrency 64. KV Offload removes the cap but pays 91~ms per step transferring the full selection. AVSG shares KV Offload's exact layout (identical device KV pool and host DRAM offload), so the gap isolates the access strategy: reusing buffer hits and transferring only misses reuse cuts TPOT 39\% to 55~ms and raises throughput $1.27\times$ to 571 tok/s, at the cost of ${\sim}4$~GB more HBM. 

\section{Conclusion}
\label{sec:conclusion}
AVSG reduces the decoding overhead of KV offloading for DSA while preserving exact token selections, showing that the overlap in selected KV entries across decoding steps is a first-class resource for serving efficiency. Its three mechanisms each remove one bottleneck: vectorized hash matching accelerates lookup $2.80\times$ over scalar dual-pointer matching, balanced miss-only transfer raises effective H2D bandwidth by up to $34.44\times$ over request-level assignment, and lifetime-based residency attains a $94.83\%$ hit rate on real requests, with shared slot reservations extending reuse across MTP tokens ($61.87\%$ even from an empty buffer under MTP-3). On a serving stack processing a real-world production dataset, AVSG cuts TPOT by $39\%$ and sustains $1.27\times$ the output throughput of the same offload layout without reuse. Matrix-level vectorization, latency-hiding state updates, and multi-tier memory hierarchies remain open directions (Appendix~\ref{app:future-work}).

\bibliography{main}
\bibliographystyle{iclr2027_conference}

\clearpage
\appendix

\section{Max Lifetime Ablation}
\label{app:lifetime}
\begin{figure}[H]
    \centering
    \begin{subfigure}[t]{0.48\textwidth}
        \centering
        \includegraphics[width=\linewidth]
        {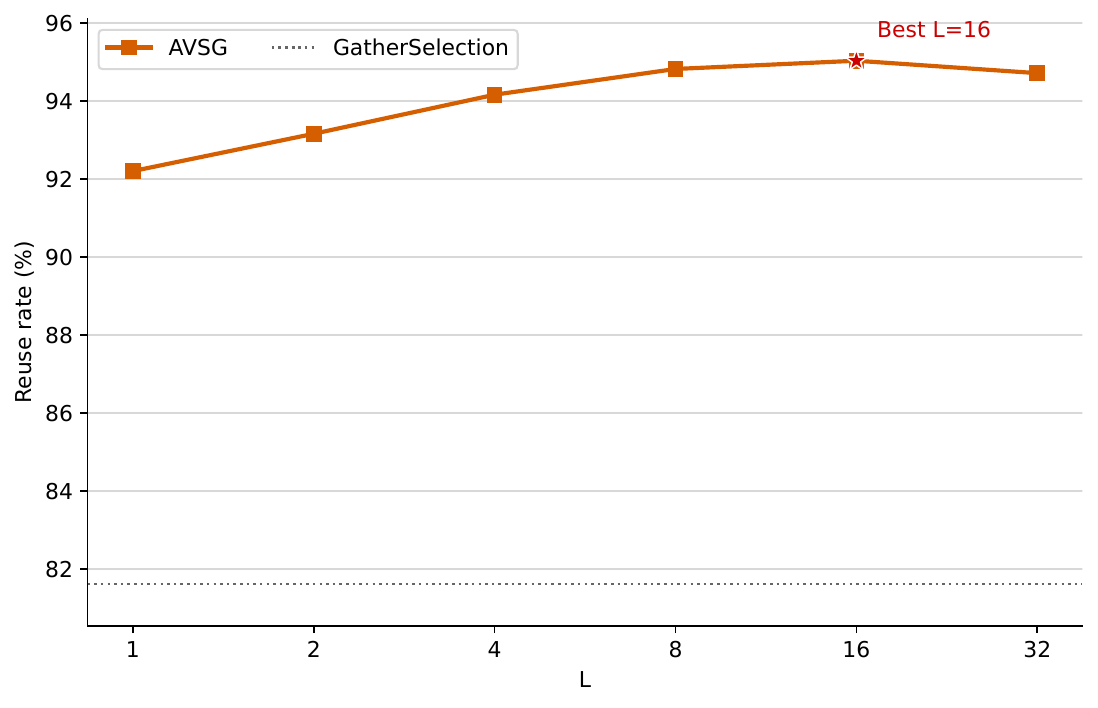}
        \caption{Reuse rate.}
        \label{fig:lifetime-reuse}
    \end{subfigure}
    \hfill
    \begin{subfigure}[t]{0.48\textwidth}
        \centering
        \includegraphics[width=\linewidth]
        {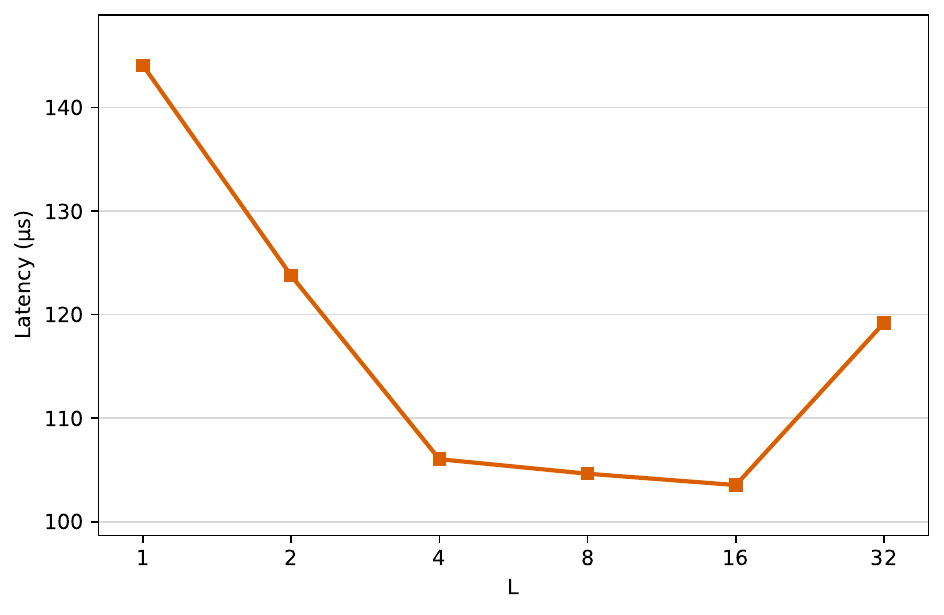}
        \caption{Mean kernel latency.}
        \label{fig:lifetime-latency-2048}
    \end{subfigure}
    \caption{Sensitivity of AVSG to the maximum record lifetime $L$
    using the 2000 real requests of \S\ref{subsec:e2e} with Top-$K=2048$,
    an 8K-entry selection pool, and MTP-0.}
    \label{fig:lifetime-ablation}
\end{figure}

\textbf{AVSG Lifetime Ablation.}
We evaluate AVSG record lifetime using traces from a representative transformer layer for the same 2000 real chat requests as \S\ref{subsec:e2e}, with Top-$K=2048$, an 8K-entry selection pool, and MTP-0. We sweep $L\in\{1,2,4,8,16,32\}$ over the same set of steady decoding steps from these requests. As shown in Figures~\ref{fig:lifetime-reuse} and~\ref{fig:lifetime-latency-2048}, the request-averaged reuse rate increases from 92.21\% at $L=1$ to 94.83\% at $L=8$ and peaks at 95.04\% for $L=16$, compared with 81.61\% for GatherSelection. Meanwhile, the mean kernel latency decreases from 104.62~$\mu$s to its minimum of 101.27~$\mu$s at $L=16$. Both metrics regress slightly at $L=32$. Therefore, we select $L=16$ as the default lifetime. 

\section{Consolidated Algorithm Details}
\label{app:alg-details}
This appendix gathers the three algorithms referenced but omitted from the main text: the vectorized state update that installs misses into the hash table, the per-step life-cycle update that consolidates the four lifetime operations of Algorithm~\ref{alg:avsg-pipeline}, and the MTP-aware processing that sequences Algorithm~\ref{alg:avsg-pipeline} over the initial token and all speculative tokens of one decode step.

\paragraph{Probe: hit masks and bucket indices.}
Given a token's selected indices $\vx=(x_0,\ldots,x_{K-1})$ and the request's hash state table $\vs$, the read-only primitive returns
\begin{equation}
(\vm,\vh)=\mathrm{Probe}(\vx,\vs),\qquad
\vm=(m_0,\ldots,m_{K-1}),\quad
\vh=(h_0,\ldots,h_{K-1}),
\label{eq:probe-outputs}
\end{equation}
where $m_i\in\{0,1\}$ indicates whether probing finds a matching selected index, and $h_i\in\{0,\ldots,S-1\}$ is the matching bucket on a hit or the final examined bucket on a miss. In particular, $h_i$ is a hash bucket index, not the destination slot $o_i$.

For each $i$, initialize the bucket and odd probe stride as
\begin{equation}
h_i^{(0)}=(a_1x_i)\bmod S,\qquad
d_i=((a_2x_i)\bmod S)\mathbin{|}1,
\label{eq:probe-initialization}
\end{equation}
where $|$ denotes bitwise OR. The superscript $r$ denotes the probe round. At each round $r=0,\ldots,R$, decode the candidate's slot field and compare its stored selected index:
\begin{equation}
o_i^{(r)}=q_{h_i^{(r)}}\bmod S,\qquad
m_i^{(r)}=\bigl[c_{o_i^{(r)}}=x_i\bigr].
\label{eq:probe-comparison}
\end{equation}
Here $o_i^{(r)}$ is the candidate slot, not an allocated destination for a miss. Square brackets around a proposition denote its Boolean indicator. For $r<R$, only unresolved probes advance:
\begin{equation}
h_i^{(r+1)}=
\begin{cases}
h_i^{(r)}, & m_i^{(r)}=1,\\
(h_i^{(r)}+d_i)\bmod S, & m_i^{(r)}=0.
\end{cases}
\label{eq:probe-advance}
\end{equation}
Since Probe does not modify the table, a matched probe remains matched. The returned values are $m_i=m_i^{(R)}$ and $h_i=h_i^{(R)}$, requiring at most $R+1$ bucket comparisons per selected index. On a hit, the caller recovers $o_i=q_{h_i}\bmod S$; on a miss, $h_i$ does not identify a reusable KV entry, and allocation supplies $o_i$ separately. Hit detection compares the full selected index; the invalid and fingerprint fields govern allocation rather than this comparison. All recurrences apply element-wise across the $K$ probes.

\begin{algorithm}[H] \small
\caption{UpdateState (vectorized)}
\label{alg:avsg-UpdateState}
\begin{algorithmic}[1]
\REQUIRE $\vt$: selected indices that miss;\quad $\vo^{\mathrm{miss}}$: allocated cache positions for $\vt$;\quad $\vs=(E_0,E_1,\ldots,E_{S-1})$: state table
\ENSURE updated $\vs$; all misses retain their allocated output positions
\STATE $\vh \gets (a_1 \cdot \vt) \bmod S$ \COMMENT{initial hash buckets}
\STATE $\vd \gets ((a_2 \cdot \vt) \bmod S)\ |\ 1$ \COMMENT{odd probe strides}
\FOR{$i = 0$ \textbf{to} $R$}
    \STATE $\vv \gets \operatorname{reusable}(q_{\vh})$ \COMMENT{Eq.~\ref{eq:reusable-hash-bucket}}
    \IF{$i < R$}
        \STATE $\vh[\neg \vv] \gets (\vh[\neg \vv] + \vd[\neg \vv]) \bmod S$ \COMMENT{advance only unresolved probes}
    \ENDIF
\ENDFOR
\STATE $E_{\vo^{\mathrm{miss}}}.c \gets \vt$ \COMMENT{all misses have allocated slots}
\STATE $q_{\vo^{\mathrm{miss}}}.\text{lifetime} \gets L$
\STATE $\widehat{\boldsymbol{\phi}} \gets \widehat{\phi}(\vt)$ \COMMENT{encoded fingerprints of selected indices that miss}
\STATE $E_{\vh[\vv]}.q \gets 2FS\,q_{\vh[\vv]}.\text{lifetime} + \widehat{\boldsymbol{\phi}}[\vv] + \vo^{\mathrm{miss}}[\vv]$
\STATE \COMMENT{One packed mapping per bucket: preserve lifetime, update fingerprint/slot, clear invalid}
\STATE \COMMENT{Same-bucket writes retain one nondeterministic writer; unresolved probes install no mapping}
\end{algorithmic}
\end{algorithm}

\textbf{Collision and probe-budget semantics.} The last assignment is a packed-word scatter. For competing writes to the same bucket, the surviving word must contain one writer's complete fingerprint and destination position, rather than fields from different writers. This expresses the hash write-collision behavior without a separate deterministic winner-selection step. The final mask $\vv$ selects probes that found a reusable bucket within rounds $0$--$R$; no write is issued for exhausted probes. Misses whose mappings are overwritten or never installed still keep their allocated slots and are copied for the current operation. Subsequent probes may miss these entries and fetch them again; a hit is accepted only after comparing the full cached token id, so losing a mapping does not create a false hit. The fingerprint belongs to the hash bucket $E_h.q$, while the cached id and its lifetime belong to the allocated slot $E_{o^{\mathrm{miss}}}$, as defined in \S\ref{subsec:hash-matching}.

\paragraph{CollectDead.}
Given the hash state table $\vs=(E_0,\ldots,E_{S-1})$, $\mathrm{CollectDead}(\vs)$ returns the indices of slots whose lifetime has expired:
\begin{equation}
\ve=\mathrm{CollectDead}(\vs)
=\bigl[j\mid 0\le j<S,\ q_j.\mathrm{lifetime}<0\bigr],
\label{eq:collect-dead}
\end{equation}
where the brackets denote a compact vector of slot indices in increasing order, each included once. The function compares all slot lifetimes with zero and compacts the matching indices without modifying the table. Here, $n_g$ denotes the number of selected indices that miss for the current token $g$. Thus, $|\ve|$ is the number of reclaimable slots, and $\ve[:n_g]$ selects the first $n_g$ slots once $|\ve|\ge n_g$. A slot with lifetime zero is not yet reclaimable; further decay and state updates are performed by the caller.

\begin{algorithm}[H] \small
\caption{Life-Cycle Update (per token)}
\label{alg:avsg-lifecycle}
\begin{algorithmic}[1]
\REQUIRE $\vs=(E_0,E_1,\ldots,E_{S-1})$: state table;\quad $L$: maximum record lifetime;\quad $\vo=(o_0,\ldots,o_{K-1})$: slots corresponding to the current token's $K$ selected indices, with hit positions populated;\quad $g$: token index within the request;\quad $\vm=(m_0,\ldots,m_{K-1})$: hit mask aligned with those selected indices;\quad $\vo^{\mathrm{miss}}$: newly allocated slots
\IF{$g=0$}
    \STATE $\vs.\text{lifetime} \gets \vs.\text{lifetime} - 1$ \COMMENT{Step 1: Global pre-decay}
\ENDIF
\STATE $q_{\vo[\vm]}.\text{lifetime} \gets L$ \COMMENT{Step 2: Refresh hits}
\STATE \COMMENT{Step 3: Ensure $\geq|\vo^{\mathrm{miss}}|$ dead slots}
\STATE $\ve \gets \mathrm{CollectDead}(\vs)$;\quad \textbf{while} $|\ve| < |\vo^{\mathrm{miss}}|$: $\vs.\text{lifetime} \gets \vs.\text{lifetime}-1$;\; $\ve \gets \mathrm{CollectDead}(\vs)$
\STATE \COMMENT{Step 4: Initialize newly allocated slots}
\STATE $q_{\vo^{\mathrm{miss}}}.\text{lifetime} \gets L$
\end{algorithmic}
\end{algorithm}

\textbf{Slot preservation under sequential token processing.}
Within each request, tokens execute Passes 1--3 sequentially. Global pre-decay runs only for $g=0$, leaving every old slot with lifetime at most $L-1$. Each token refreshes its hits and initializes its newly allocated slots to $L$ before the next token is processed; slots selected only by later tokens are not refreshed in advance.

To bound reclamation across the request, consider the cumulative unit decays performed during allocation. After $L$ such decays, every old slot that has never been refreshed is reclaimable. Any slot refreshed or initialized by a processed token still has nonnegative lifetime, since it has undergone at most $L$ decays since being set to $L$. Before allocating the current token's misses, the number of slots already required by processed selections plus those misses is at most $N_qK\le S$. Thus, by the $L$th cumulative decay, enough other slots are reclaimable, and no further decay is needed. All output slots already assigned within the request therefore remain available until Stage Two and attention consume them. An old entry selected only by a later token may be reclaimed earlier; that token then treats it as a miss and fetches it again. This policy provides lifetime-based aging rather than exact LRU replacement.
\begin{algorithm}[H] \small
\caption{MTP-Aware AVSG Processing (per decode step)}
\label{alg:avsg-mtp}
\begin{algorithmic}[1]
\REQUIRE $\mX\in\mathbb{Z}^{(N_{\mathrm{mtp}}+1)\times K}$: top-$K$ index matrix;\quad $\vs=(E_0,E_1,\ldots,E_{S-1})$: shared state table;\quad $N$: verified context length (excluding speculative MTP tokens)
\ENSURE $\mO\in\mathbb{Z}^{(N_{\mathrm{mtp}}+1)\times K}$: buffer position matrix;\quad updated $\vs$
\STATE \COMMENT{Stage One: execute all three passes for each token in sequence}
\FOR{$g = 0$ \textbf{to} $N_q-1$}
    \STATE Set $\vx$ to row $g$ of $\mX$ and execute Algorithm~\ref{alg:avsg-pipeline} with token index $g$ and table $\vs$
    \STATE Store $\vo$ in row $g$ of $\mO$ and append this token's misses and destination slots to the transfer list
\ENDFOR
\STATE Synchronize all matching cores (\texttt{SyncAll})
\STATE Execute Stage Two for the accumulated miss list (\S\ref{subsec:two-stage})
\STATE Complete H2D transfers before attention consumes $\mO$
\STATE \COMMENT{End of step: speculative invalidation}
\STATE $\vv \gets \bigl(\vs.c \ge N\bigr)$
\STATE $E_{\vv}.c \gets {-}1$ \COMMENT{mark speculative entries as empty}
\end{algorithmic}
\end{algorithm}

\textbf{Speculative Invalidation.}
The $N_{\mathrm{mtp}}$ additional tokens are speculative: only those passing verification are committed, while rejected tokens trigger KV regeneration on the host in the next step.
Retaining entries for rejected tokens would make hash matching return spurious hits---reading stale device kv instead of the regenerated ones, causing \emph{silent accuracy degradation}.
The end-of-step invalidation in Algorithm~\ref{alg:avsg-mtp} prevents this by marking every entry whose cached id $c_j\ge N$ (the speculative range beyond the verified context length) as empty, forcing misses and fresh H2D transfers of the regenerated data.
Invalidating \emph{all} speculative entries---including those that may pass verification---costs at most $N_{\mathrm{mtp}}\times K$ extra miss lookups in the next step, negligible against the correctness guarantee.

\section{Vectorized Hash Matching: Full Protocol and Results}
\label{app:hashmatch-details}

We isolate Stage One and compare the vectorized open-addressing hash matching against the scalar dual-pointer implementation it replaces. Both variants run in an identical test harness and differ only in the matching algorithm, so any latency difference is attributable to the algorithm alone. We report the median \texttt{Duration} of the algorithm over 1000 consecutive iterations, collected with the Profiler Tools on a single NPU; median, trimmed mean, and arithmetic mean agree to within 2\% at every operating point, so no conclusion below depends on the choice of statistic.

\textbf{Matching Performance.} Under a matched configuration in which both implementations face an identical working set, vectorized matching overtakes the scalar dual-pointer just above $N{=}512$ and is $2.80\times$ faster at $N{=}2048$ (Figure~\ref{fig:hashmatch}(a)). The matched configuration is dictated by the baseline: the dual-pointer implementation couples its buffer capacity to the top-K length, admitting only $S = K$ and exposing no independent control over~$S$. We therefore sweep $S = K = N$ for both variants---conceding nothing to the vectorized algorithm, whose cost is nearly independent of~$S$ (Figure~\ref{fig:hashmatch}(c)), while removing the one degree of freedom the baseline cannot exercise. Under this protocol the scalar cost grows linearly in the number of indices to match, from $20.8\,\mu$s at $N{=}256$ to $150.0\,\mu$s at $N{=}2048$, whereas over the same $8\times$ range the vectorized cost moves only from $38.2$ to $53.7\,\mu$s. Below the crossover the scalar path is cheaper (the vectorized implementation carries a larger fixed overhead); above it the linear term dominates and the gap widens monotonically.

\textbf{Decoupling from $K$ and $S$.} With~$S$ fixed at 4096 entries, an $8\times$ increase in~$K$ costs only $1.27\times$ more latency, from $46.0$ to $58.6\,\mu$s (Figure~\ref{fig:hashmatch}(b))---the same $8\times$ increase raises the scalar cost by $7.2\times$; with~$K$ fixed at 512, a $16\times$ increase in~$S$, from 512 to 8192, costs $1.60\times$, from $38.6$ to $61.9\,\mu$s (Figure~\ref{fig:hashmatch}(c)). Because the hash probe is $O(1)$ SIMD per entry, lookup cost is governed by the retry budget~$R$ rather than by table capacity---this is the property AVSG depends on: the buffer can be enlarged to raise the hit rate and cut H2D traffic without a proportional matching penalty. The residual drift (larger $K$ or $S$ presents a longer vector to sweep, and added elements are not entirely free) matters only in ratio: latency rises roughly an order of magnitude more slowly than the volume of vector data matched, whereas the scalar path pays $7.2\times$ for an $8\times$ larger working set---close to proportional. The $S$ dimension has no scalar counterpart to plot, since the dual-pointer admits only $S = K$; a baseline pursuing the same hit-rate benefit would have to raise~$K$ in lockstep and pay the linear cost of Figure~\ref{fig:hashmatch}(a).

\section{H2D Core-Binding Strategies: Definitions and Full Results}

\label{app:h2d-details}

This appendix defines the three core-binding strategies compared in \S\ref{subsec:h2d-efficiency}, analyzes their transfer workloads, and presents the experimental setup and full bandwidth results.

\textbf{S1 (Request-level)} assigns one core to each request, processing all miss KV entries within that request serially; parallelism is capped by the batch size $B$, leaving most cores idle when $B$ is small.

\textbf{S2 (Top-K position level)} pools all top-K positions from all top-K index sets into a single array and splits it evenly across all $C$ cores by fixed-width partition, so each core receives an equal share of positions regardless of their hit/miss status; cores issue transfers for every assigned position, with no-op transfers for hit positions wasting transfer capacity proportional to the cache hit rate.

\textbf{S3 (Miss-KVs-level, AVSG)} pools all cache misses into a single miss list and distributes them evenly across $C$ cores (\S\ref{subsec:two-stage} of the main text); every issued transfer corresponds to a real miss KV, so no transfer capacity is wasted on hits.

\textbf{Transfer imbalance analysis.} Using the notation of \S\ref{subsec:two-stage} and the workload definition in Eq.~\ref{eq:core-transfer-workload}, we analyze why request-level assignment and equal top-$K$ partitioning can produce imbalanced H2D transfers.

Under request-level assignment, distribute the $B$ requests as evenly as possible across the $C$ cores, assigning one extra request to each of the first $r_B=B\bmod C$ cores. Each request contributes $N_qK$ selected-index positions, so
\begin{equation}
|\mathcal A_c^{\mathrm{req}}|=
\begin{cases}
\bigl(\lfloor B/C\rfloor+1\bigr)N_qK, & 0\le c<r_B,\\
\lfloor B/C\rfloor N_qK, & r_B\le c<C.
\end{cases}
\label{eq:request-assignment}
\end{equation}
When $B$ is not divisible by $C$, $|\mathcal A_c^{\mathrm{req}}|$ varies across cores, which can lead to imbalanced transfer workloads $W_c$.
To balance assigned positions at a finer granularity, equal top-$K$ partitioning flattens all $N_gK$ positions and divides them into near-equal contiguous ranges:
\begin{equation}
\mathcal A_c^{\mathrm{pos}}
=\left\{(g,i):
\left\lfloor\frac{cN_gK}{C}\right\rfloor
\le gK+i <
\left\lfloor\frac{(c+1)N_gK}{C}\right\rfloor
\right\},
\label{eq:position-assignment}
\end{equation}
where $0\le g<N_g$ and $0\le i<K$. These sets differ in size by at most one, but cache-hit counts may still differ across cores. For two distinct cores $c,d\in\{0,\ldots,C-1\}$ with $|\mathcal A_c^{\mathrm{pos}}|=|\mathcal A_d^{\mathrm{pos}}|$, their transfer-workload difference is
\begin{equation}
W_c-W_d
=\sum_{(g,i)\in\mathcal A_d^{\mathrm{pos}}}m_{g,i}
-\sum_{(g,i)\in\mathcal A_c^{\mathrm{pos}}}m_{g,i}.
\label{eq:hit-induced-imbalance}
\end{equation}
Thus, differences in cache-hit counts directly produce transfer imbalance; a partition containing only hits has $W_c=0$ even if another partition still requires transfers.

\textbf{Experimental setup.} To evaluate these strategies, we select batch sizes $B\in\{1,2,4,8,16,32\}$ from the source CSV and cache reuse rates $r\in\{0,0.25,0.5,0.75,1\}$, with miss rate $m=1-r$, $K=2048$, $N_{\mathrm{mtp}}=0$ (one initial token per request, no speculative tokens), $H=1$, and 1152 bytes per selected token.

\textbf{Metric.} For each (strategy, batch size, miss rate) combination, we compute the effective H2D bandwidth in GB/s. Only miss cache are transferred, so the number of moved tokens is $B\,(N_{\mathrm{mtp}}+1)\,H\,K\,m$, and at $1152$ bytes per token the transfer volume is $V = 1152\,B\,(N_{\mathrm{mtp}}+1)\,H\,K\,m$ bytes. Dividing $V$ by the measured wall-clock transfer time $t$ yields the effective bandwidth $V/t$.

\textbf{Speedup source data.} Table~\ref{tab:h2d-full-speedup} reports the CSV columns \texttt{BW\_S1\_baseline\_GBps}, \texttt{BW\_S2\_master\_GBps}, and \texttt{BW\_S3\_balance\_GBps}, with speedups computed as S3/S1 and S3/S2 before rounding. For every nonzero transfer volume this is also the inverse ratio of transfer times. At 100\% reuse the transfer volume is zero; kernel runtimes may remain nonzero, but they do not define a bandwidth ratio.

\begin{figure}[t]
    \centering
    \includegraphics[width=\linewidth]{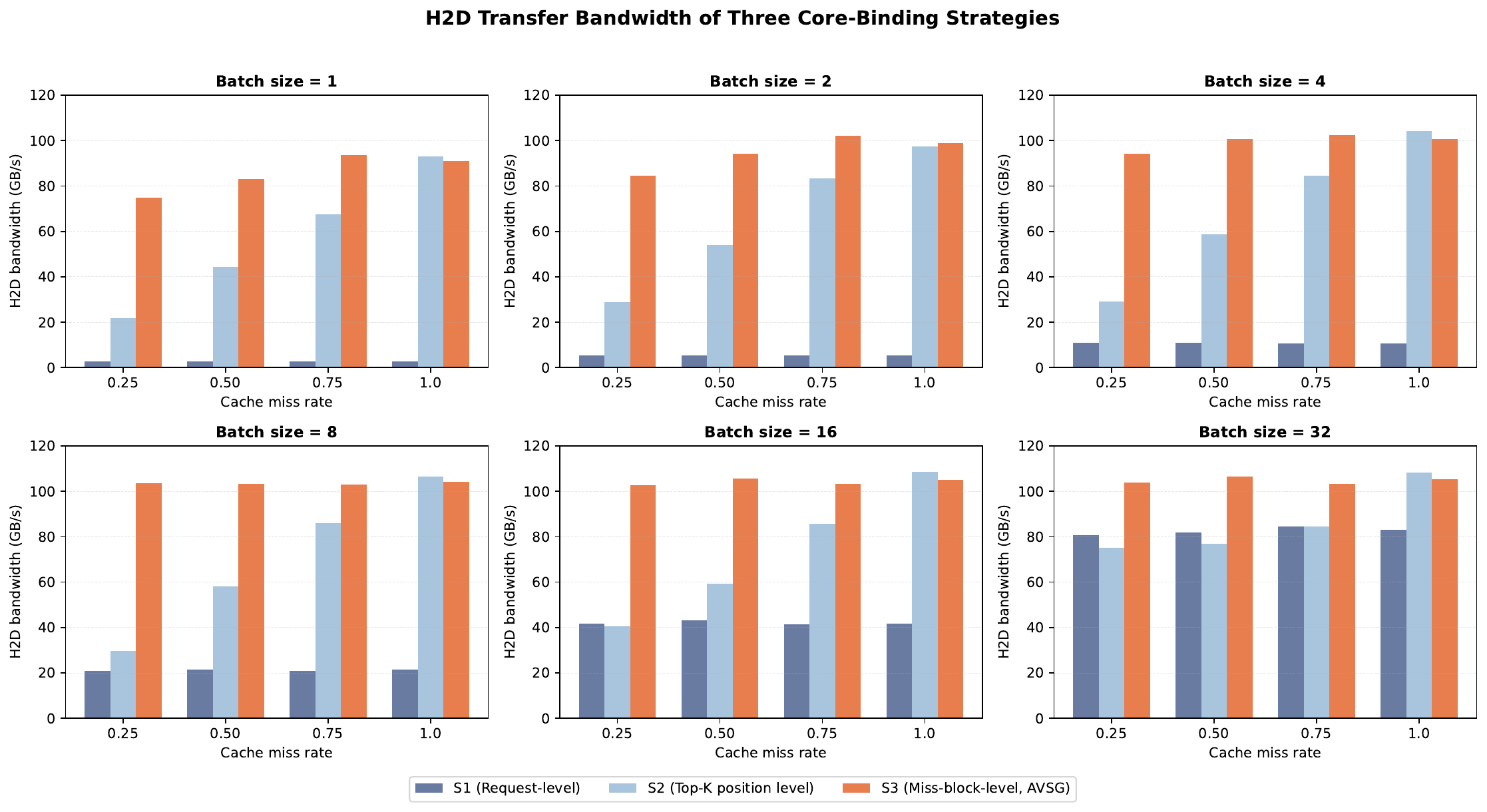}
    \caption{{Effective H2D bandwidth (GB/s) of three core-binding strategies under varying batch sizes ($B=1,2,4,8,16,32$) and cache miss rates ($m=0.25,0.5,0.75,1.0$). Each subplot corresponds to one batch size. Within each subplot, the three bars per miss rate represent S1 (request-level binding), S2 (Top-K position level binding), and S3 (miss-KV-level binding, AVSG). Higher bandwidth indicates more efficient use of the transfer engines. Experimental configuration: top-$K=2048$, $N_{\mathrm{mtp}}=0$, $H=1$, per-token payload $(64+512)\times2=1152$ bytes.}}
    \label{fig:stagetwo-bandwidth}
\end{figure}

\textbf{Full results.} Figure~\ref{fig:stagetwo-bandwidth} compares absolute bandwidth across the tested batch sizes and nonzero miss rates. Table~\ref{tab:h2d-full-speedup} provides the full measurements and speedup ratios, including the zero-transfer case at 100\% reuse.

\begin{table}[h]
\centering
\footnotesize
\caption{S1, S2, and S3 bandwidth comparison for batch sizes 1 to 32. Bandwidth values retain the source precision (GB/s); speedups are S3/S1 and S3/S2, rounded to two decimals. The last row covers all six batch sizes at 100\% reuse: no KV data are transferred, so bandwidth speedup is undefined.}
\label{tab:h2d-full-speedup} \small
\begin{tabular}{@{}rrrrrrr@{}}
\toprule
Batch size & Reuse (\%) & S1 (GB/s) & S2 (GB/s) & S3 (GB/s) & S3/S1 & S3/S2 \\
\midrule
1 & 0 & 2.7164 & 93.0307 & 90.8505 & $33.45\times$ & $0.98\times$ \\
1 & 25 & 2.7205 & 67.6668 & 93.7022 & $34.44\times$ & $1.38\times$ \\
1 & 50 & 2.7351 & 44.4206 & 83.1734 & $30.41\times$ & $1.87\times$ \\
1 & 75 & 2.6629 & 21.9380 & 74.7369 & $28.07\times$ & $3.41\times$ \\
\addlinespace[2pt]
2 & 0 & 5.3596 & 97.4847 & 98.8166 & $18.44\times$ & $1.01\times$ \\
2 & 25 & 5.4038 & 83.2732 & 102.2639 & $18.92\times$ & $1.23\times$ \\
2 & 50 & 5.3269 & 53.9419 & 94.1647 & $17.68\times$ & $1.75\times$ \\
2 & 75 & 5.2613 & 28.7019 & 84.4718 & $16.06\times$ & $2.94\times$ \\
\addlinespace[2pt]
4 & 0 & 10.6228 & 104.2751 & 100.6043 & $9.47\times$ & $0.96\times$ \\
4 & 25 & 10.6920 & 84.4470 & 102.3319 & $9.57\times$ & $1.21\times$ \\
4 & 50 & 10.8849 & 58.7102 & 100.6075 & $9.24\times$ & $1.71\times$ \\
4 & 75 & 10.8392 & 29.1469 & 94.0896 & $8.68\times$ & $3.23\times$ \\
\addlinespace[2pt]
8 & 0 & 21.4230 & 106.6558 & 104.2904 & $4.87\times$ & $0.98\times$ \\
8 & 25 & 20.8603 & 85.8649 & 102.9668 & $4.94\times$ & $1.20\times$ \\
8 & 50 & 21.5623 & 58.0890 & 103.2277 & $4.79\times$ & $1.78\times$ \\
8 & 75 & 20.9256 & 29.8066 & 103.4507 & $4.94\times$ & $3.47\times$ \\
\addlinespace[2pt]
16 & 0 & 41.6336 & 108.4312 & 104.9474 & $2.52\times$ & $0.97\times$ \\
16 & 25 & 41.4463 & 85.8342 & 103.2037 & $2.49\times$ & $1.20\times$ \\
16 & 50 & 43.0587 & 59.2252 & 105.5702 & $2.45\times$ & $1.78\times$ \\
16 & 75 & 41.6196 & 40.4595 & 102.7434 & $2.47\times$ & $2.54\times$ \\
\addlinespace[2pt]
32 & 0 & 83.1604 & 108.4119 & 105.4712 & $1.27\times$ & $0.97\times$ \\
32 & 25 & 84.4773 & 84.4850 & 103.3840 & $1.22\times$ & $1.22\times$ \\
32 & 50 & 81.7463 & 76.8601 & 106.5611 & $1.30\times$ & $1.39\times$ \\
32 & 75 & 80.8333 & 75.0018 & 103.9796 & $1.29\times$ & $1.39\times$ \\
\addlinespace[2pt]
All & 100 & 0.0000 & 0.0000 & 0.0000 & -- & -- \\
\bottomrule
\vspace{-5mm}
\end{tabular}
\end{table}

\textbf{Bandwidth trends.} S1's effective bandwidth scales almost linearly with batch size---from $2.7$~GB/s at $B{=}1$ to ${\sim}83$~GB/s at $B{=}32$---because its assignment granularity is the request, so parallelism cannot exceed $B$; even at $B{=}32$ it trails S3 by $20$--$25$~GB/s since misses within each request remain serialized. S2 matches S3 when $m{=}1.0$ (every position is a genuine miss; $108$ vs.\ $104.9$~GB/s at $B{=}16$, the small excess reflecting its simpler equal-width partition), but as $m$ decreases its cores issue no-op transfers for hit positions: at $B{=}16$, $m{=}0.25$, S2 drops to $40.5$~GB/s while S3 holds $102.7$~GB/s. S3's dynamic work partition ensures every issued transfer points to a real miss KV, and the work is spread evenly across all $C$ engines regardless of batch size or miss rate.

\textbf{Relation to the main-text comparison.} Figure~\ref{fig:h2d-speedup} uses S3/S1 on a logarithmic vertical axis and S3/S2 on a linear vertical axis; both show categorical batch sizes 1--32 and reuse rates of 0--75\%, with a dashed line at equal bandwidth. The largest gains are $34.44\times$ at $B=1$, $r=0.25$ against request-level assignment and $3.48\times$ at $B=8$, $r=0.75$ against equal top-$K$ partitions. Equal numbers of assigned positions do not guarantee equal numbers of misses: cache hits can leave different amounts of useful transfer work on different cores, while hit positions also incur no-op issue overhead in this baseline. AVSG instead partitions the actual miss list. The bandwidth measurements capture the combined effect of these costs, rather than isolating their individual contributions. At zero reuse, equal-position and miss-only partitioning have similar workloads and near-parity bandwidth; the full values are reported in Table~\ref{tab:h2d-full-speedup}.

\section{Cache Reuse: Capacity and MTP Results}
\label{app:reuse-details}

This appendix reports the full synthetic protocol behind the selection-pool capacity study of \S\ref{subsec:cache-reuse}. The synthetic experiments use a maximum KV length of 16K and five random seeds, with Top-$K{=}2048$, lifetime $L{=}8$, and MTP-0. We fix the adjacent-step similarity $\gamma$ to $0.5$, $0.7$, or $0.9$, while sampling historical recall from $\mathcal{N}(\mu_{\mathrm{hist}}, 0.01^2)$ with $\mu_{\mathrm{hist}}\in\{0.5,0.7,0.9\}$, clipped to $[0,1]$. Table~\ref{tab:fixed-adjacent-history-normal} lists the reuse rate of the 4K and 8K pools in all nine settings and on the 2000 real chat requests used in \S\ref{subsec:e2e}; the improvement is larger under low-to-moderate locality, where additional capacity preserves more reusable entries from earlier decoding steps, and gradually diminishes as the reuse rate approaches saturation.

\begin{table}[t]
  \centering
  \small
  \caption{AVSG reuse rate (\%) for Top-$K{=}2048$ with 4K and 8K selection pools. Synthetic: max KV length 16K, $L{=}8$, MTP-0, five seeds; adjacent-step similarity $\gamma$ fixed, historical recall $\sim\mathcal{N}(\mu_{\mathrm{hist}}, 0.01^2)$, clipped to $[0,1]$. Real: request-level macro over the 2000 requests used in \S\ref{subsec:e2e}.}
  \label{tab:fixed-adjacent-history-normal}
  \begin{tabular}{lccrr}
    \toprule
    Data & $\gamma$ & $\mu_{\mathrm{hist}}$ & 4K pool & 8K pool \\
    \midrule
    Synthetic & 50.00\% & 0.5 & 60.00\% & 68.19\% \\
    Synthetic & 50.00\% & 0.7 & 70.44\% & 81.27\% \\
    Synthetic & 50.00\% & 0.9 & 84.29\% & 94.15\% \\
    Synthetic & 70.00\% & 0.5 & 74.76\% & 82.55\% \\
    Synthetic & 70.00\% & 0.7 & 81.94\% & 89.99\% \\
    Synthetic & 70.00\% & 0.9 & 90.12\% & 96.79\% \\
    Synthetic & 90.00\% & 0.5 & 88.90\% & 94.91\% \\
    Synthetic & 90.00\% & 0.7 & 92.61\% & 97.02\% \\
    Synthetic & 90.00\% & 0.9 & 96.21\% & 99.04\% \\
    \midrule
    Real & 81.61\% & -- & 88.72\% & 94.83\% \\
    \bottomrule
  \end{tabular}
\end{table}

\textbf{Selection of main-text configurations.} For each fixed adjacent-step reuse $\gamma$, we select the historical recall setting that maximizes the 8K-buffer hit rate minus $100\gamma$ percentage points. Among the three tested historical recall means, $\mu_{\mathrm{hist}}=0.9$ maximizes this gain at each $\gamma\in\{0.5,0.7,0.9\}$, giving gains of 44.15, 26.79, and 9.04 pp, respectively. These are best-case configurations within the sweep, not averages across historical recall settings.

\begin{figure}[H]
  \centering
  \includegraphics[width=0.6\linewidth]
  {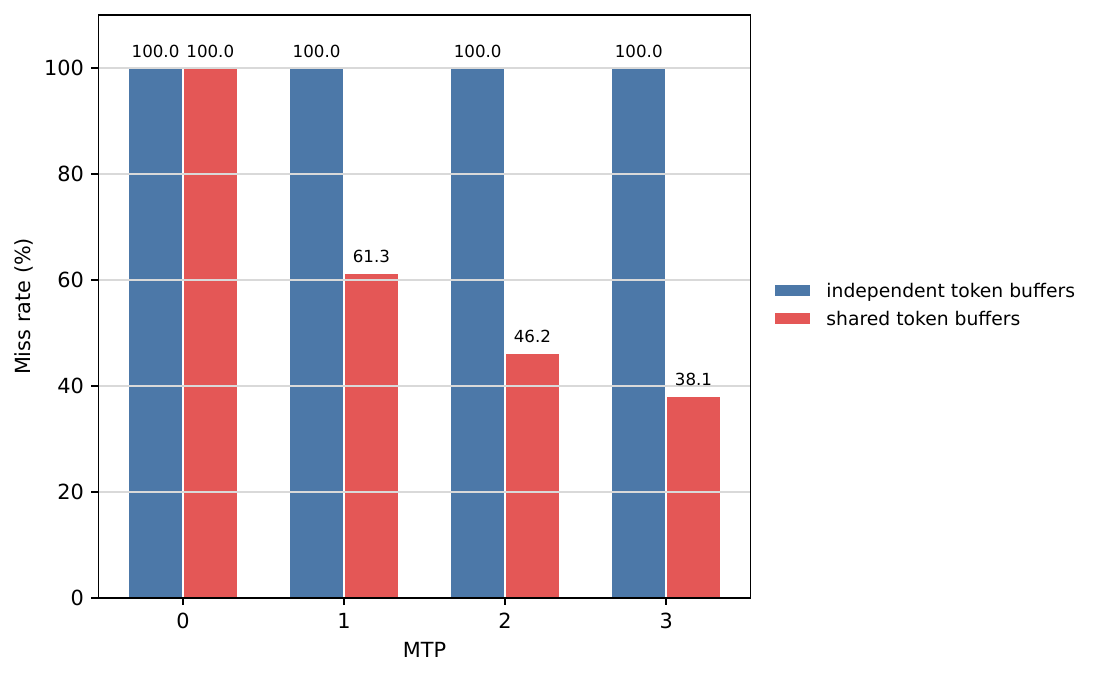}
  \caption{Cold-start cache miss rate with shared and independent token buffers. Sharing the cache buffer across MTP tokens enables intra-iteration reuse and progressively reduces the miss rate as the MTP degree increases.}
  \label{fig:mtp-affinity}
\end{figure}

\textbf{Cold-start MTP protocol.} On the same 2000 real chat requests used in \S\ref{subsec:e2e}, we fix $K=2048$ and $S=8$K and simulate MTP degrees 0--3, where degree $d=N_{\mathrm{mtp}}$ includes the initial decode token and $d$ additional speculative tokens. All speculative tokens are assumed accepted. Each request starts with an empty cache, and reuse is measured over all selected KV entries in its first decode iteration, then averaged across requests. Independent processing starts each token with an empty buffer and yields zero reuse. Shared-buffer processing records the first token's misses and their reserved slots for subsequent tokens, then transfers all misses after matching; the first token itself still has zero reuse. The resulting request-averaged reuse rates are 0\%, 38.71\%, 53.79\%, and 61.87\% for degrees 0--3, respectively (Figure~\ref{fig:mtp-affinity}).

\section{Metadata-Word Decode Chain}
\label{app:decode-chain}
This appendix expands the inline decode of \S\ref{subsec:stage-one} into its full equation form. For the packed metadata word $q_h$ stored in bucket $h$ (Eq.~\ref{eq:packed-hash-mapping} of the main text), AVSG first extracts the index field:
\begin{equation}
u_h = q_h \bmod (2FS).
\end{equation}
The invalid flag, mapped cache position, and stored fingerprint are then decoded as
\begin{align}
b_h &= \left\lfloor \frac{u_h}{FS} \right\rfloor, &
o_h &= u_h \bmod S, &
\phi_h &= \left\lfloor \frac{u_h \bmod (FS)}{S} \right\rfloor,
\label{eq:decode-hash-mapping}
\end{align}
and since $0\le u_h<2FS$, the decoded invalid flag satisfies $b_h\in\{0,1\}$. A bucket contains a valid mapping iff
\begin{equation}
\operatorname{valid}(q_h)=[b_h=0],
\end{equation}
which feeds the reusable predicate of Eq.~\ref{eq:reusable-hash-bucket} in the main text.


\section{End-to-End Serving: Full Experimental Setting}
\label{app:e2e-details}

\textbf{Serving platform.} All settings run on a 12-node Ascend 910B NPU (64~GB HBM per chip) cluster with a 5P1D dis-aggregated topology---5 prefill instances (each 2 nodes, TP=16, EP=16) plus 1 decode instance (2 nodes, DP=16, EP=16)---serving GLM-5.3-w4a8c16 with 3-token speculative decoding. The workload is an agent dataset with a total of 2000 real chat requests collected from a real-world production system, with an average prompt length of 
48k tokens. We evaluate with a maximum output of 32768 tokens, a concurrency of 64, and a request rate of 1.  The benchmark script extends from \citep{wang2025omniinfer} to maintain a fixed concurrency or request rate throughout the evaluation, so that the sustained load drives the system toward its throughput upper bound; as throughput rises, TTFT grows accordingly. Each setting starts from a fresh server with an empty prefix cache, and all settings serve identical request streams. In the offload settings, MLA latent KV is swapped to host DRAM via hugepage mmap into a 632~GB pool, while HBM retains the DSA retrieval structure plus a device-side KV window.

\textbf{Memory layout.} Because the three settings differ in where KV lives, we verify the layout from server logs rather than assuming it. After model loading, the profile stage leaves 16.22~GB HBM available per chip under No Offload versus 19.50~GB under both offload settings (identical code path), which caps No Offload's device KV pool at 145{,}536 tokens versus 762{,}496 tokens ($5.2\times$) for offload---and caps its supported context at 145{,}536 versus 524{,}288 tokens. Crucially, AVSG and KV offload share the {identical} memory layout; the only difference between them is the access strategy (per-step full H2D versus AVSG's hit reuse with miss-only transfer), so the performance gap measured between them is attributable to gather selection alone.

\section{Future Directions}
\label{app:future-work}

\textbf{Matrix-Level Vectorization.} While AVSG currently employs 256-bit SIMD vectorization for hash matching, the inherent parallelism in sparse token lookup could be further exploited through matrix-level operations. By reformulating the matching algorithm as batched matrix computations, we could leverage tensor cores and matrix multiplication units on modern accelerators, potentially achieving even higher throughput than the current 200GB/s.

\textbf{Decoupled State Update with Latency Hiding.} Profiling reveals that most latency in Stage One matching comes from state record updates (lifetime counter maintenance for LRU eviction). Critically, these state updates do not affect the subsequent attention computation flow---they only influence future cache eviction decisions. This dependency structure enables two optimizations: (1) overlap state record updates with Stage Two data transfer or attention computation, effectively hiding the update latency, and (2) compute attention for cache-hit tokens immediately while simultaneously loading cache-miss tokens in parallel, reducing the critical path latency for mixed hit/miss scenarios.

\textbf{Additional Directions.} Other promising directions include: (1) adaptive hash table sizing based on workload characteristics and top-$K$ patterns, (2) learned prefetching policies that predict future access patterns from attention distributions, (3) cross-layer optimization with DSA's token selection to minimize transfer latency, (4) support for dynamic sparse attention patterns that change during generation, and (5) extensions to multi-tier memory hierarchies including NVMe and CXL-attached memory. These enhancements could further improve the efficiency of sparse attention serving systems beyond DeepSeek Sparse Attention.

\end{document}